\documentclass[aps,prl,showpacs,preprint,superscriptaddress]{revtex4}

\usepackage{epsfig}

\newcommand{\mpl}{M_{p\bar{\Lambda}}}
\newcommand{\mb}{{M_{\rm bc}}}
\newcommand{\de}{{\Delta{E}}}
\newcommand{\plam}{{p\bar{\Lambda}}}
\newcommand{\plpi}{{p\bar{\Lambda}\pi^-}}
\newcommand{\plg}{{p\bar{\Lambda}\gamma}}

\newcommand{\psigg}{{p\bar{\Sigma}^0\gamma}}

\begin{document}



\epsfysize 25mm \epsfbox{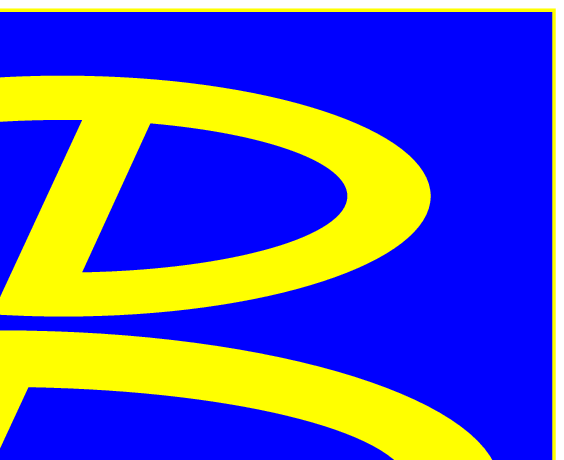}

\begin{flushright}
\vskip -25mm \noindent \hspace*{3.0in}{\bf
                    BELLE-CONF-0413
                  \\ICHEP Abstract 11-0658
}
\end{flushright}


\title{ \quad\\[1cm] \Large
Observation of $B^+ \to \plg$}

\tighten


\affiliation{Aomori University, Aomori}
\affiliation{Budker Institute of Nuclear Physics, Novosibirsk}
\affiliation{Chiba University, Chiba}
\affiliation{Chonnam National University, Kwangju}
\affiliation{Chuo University, Tokyo}
\affiliation{University of Cincinnati, Cincinnati, Ohio 45221}
\affiliation{University of Frankfurt, Frankfurt}
\affiliation{Gyeongsang National University, Chinju}
\affiliation{University of Hawaii, Honolulu, Hawaii 96822}
\affiliation{High Energy Accelerator Research Organization (KEK), Tsukuba}
\affiliation{Hiroshima Institute of Technology, Hiroshima}
\affiliation{Institute of High Energy Physics, Chinese Academy of Sciences, Beijing}
\affiliation{Institute of High Energy Physics, Vienna}
\affiliation{Institute for Theoretical and Experimental Physics, Moscow}
\affiliation{J. Stefan Institute, Ljubljana}
\affiliation{Kanagawa University, Yokohama}
\affiliation{Korea University, Seoul}
\affiliation{Kyoto University, Kyoto}
\affiliation{Kyungpook National University, Taegu}
\affiliation{Swiss Federal Institute of Technology of Lausanne, EPFL, Lausanne}
\affiliation{University of Ljubljana, Ljubljana}
\affiliation{University of Maribor, Maribor}
\affiliation{University of Melbourne, Victoria}
\affiliation{Nagoya University, Nagoya}
\affiliation{Nara Women's University, Nara}
\affiliation{National Central University, Chung-li}
\affiliation{National Kaohsiung Normal University, Kaohsiung}
\affiliation{National United University, Miao Li}
\affiliation{Department of Physics, National Taiwan University, Taipei}
\affiliation{H. Niewodniczanski Institute of Nuclear Physics, Krakow}
\affiliation{Nihon Dental College, Niigata}
\affiliation{Niigata University, Niigata}
\affiliation{Osaka City University, Osaka}
\affiliation{Osaka University, Osaka}
\affiliation{Panjab University, Chandigarh}
\affiliation{Peking University, Beijing}
\affiliation{Princeton University, Princeton, New Jersey 08545}
\affiliation{RIKEN BNL Research Center, Upton, New York 11973}
\affiliation{Saga University, Saga}
\affiliation{University of Science and Technology of China, Hefei}
\affiliation{Seoul National University, Seoul}
\affiliation{Sungkyunkwan University, Suwon}
\affiliation{University of Sydney, Sydney NSW}
\affiliation{Tata Institute of Fundamental Research, Bombay}
\affiliation{Toho University, Funabashi}
\affiliation{Tohoku Gakuin University, Tagajo}
\affiliation{Tohoku University, Sendai}
\affiliation{Department of Physics, University of Tokyo, Tokyo}
\affiliation{Tokyo Institute of Technology, Tokyo}
\affiliation{Tokyo Metropolitan University, Tokyo}
\affiliation{Tokyo University of Agriculture and Technology, Tokyo}
\affiliation{Toyama National College of Maritime Technology, Toyama}
\affiliation{University of Tsukuba, Tsukuba}
\affiliation{Utkal University, Bhubaneswer}
\affiliation{Virginia Polytechnic Institute and State University, Blacksburg, Virginia 24061}
\affiliation{Yonsei University, Seoul}
  \author{K.~Abe}\affiliation{High Energy Accelerator Research Organization (KEK), Tsukuba} 
  \author{K.~Abe}\affiliation{Tohoku Gakuin University, Tagajo} 
  \author{N.~Abe}\affiliation{Tokyo Institute of Technology, Tokyo} 
  \author{I.~Adachi}\affiliation{High Energy Accelerator Research Organization (KEK), Tsukuba} 
  \author{H.~Aihara}\affiliation{Department of Physics, University of Tokyo, Tokyo} 
  \author{M.~Akatsu}\affiliation{Nagoya University, Nagoya} 
  \author{Y.~Asano}\affiliation{University of Tsukuba, Tsukuba} 
  \author{T.~Aso}\affiliation{Toyama National College of Maritime Technology, Toyama} 
  \author{V.~Aulchenko}\affiliation{Budker Institute of Nuclear Physics, Novosibirsk} 
  \author{T.~Aushev}\affiliation{Institute for Theoretical and Experimental Physics, Moscow} 
% ...
\collaboration{The Belle Collaboration}

\begin{abstract}

We report the first observation of the radiative hyperonic $B$ decay
$B^+ \to \plg$, using a 140 fb$^{-1}$ data sample recorded on the
$\Upsilon({\rm 4S})$ resonance with the Belle detector at the KEKB
asymmetric energy $e^+e^-$ collider. The measured branching
fraction is ${\mathcal B}(B^+ \to \plg) = (2.16 ^{+0.58}_{-0.53}
\pm 0.20) \times 10^{-6}$. A search for $B^+ \to \psigg$ yields no
significant signal, so we set a 90\% confidence-level upper
limit on the branching fraction 
of ${\mathcal B}(B^+ \to \psigg) < 3.3 \times 10^{-6}$.

\vskip1pc
\pacs{ 13.40.Hq, 14.40.Nd, 14.20.Dh, 14.20.Jn}

\end{abstract}
%
%
%

\maketitle
{\renewcommand{\thefootnote}{\fnsymbol{footnote}}

\setcounter{footnote}{0}

\normalsize

%
%


The $ b \to s \gamma$ penguin diagram plays an essential role for
the large rates of the observed radiative $B \to K^*
\gamma$~\cite{kstarg} decays. It is also a good probe of new
physics beyond the Standard Model~\cite{btosg}. Recently, the Belle
collaboration reported a very stringent limit of ${\mathcal O} (10^{-6})$ on
the branching fraction of two-body $B^+ \to \plam$
decays~\cite{2body} but found an unexpectedly large rate for the
three-body decay $B^0 \to \plpi$~\cite{plpi}, which proceeds
presumably via the $b \to s$ penguin process. One interesting
feature of the $B^0 \to \plpi$ decay is that the observed
proton-$\bar{\Lambda}$ invariant mass $M_{\plam}$ spectrum peaks
near threshold. Naively, a suppression of  ${\mathcal O}(\alpha_{EM})$ is
expected for the $B^+ \to \plg$  decay relative to $B^+ \to \plam$
if the former process is bremsstrahlung-like.
In contrast, the short-distance
$ b \to s \gamma$ contribution can lead naturally
to a non-bremsstrahlung-like energetic photon spectrum and an
enhancement of  $M_{\plam}$  at low mass; the former
distribution can be compared to the recently measured $b \to
s\gamma$ inclusive photon energy spectrum~\cite{gammaspec}. These
features motivate our study of $B^+\to\plg$.
Some theoretical
predictions~\cite{theory} for the branching fraction of $B^+ \to
\plg$ are at the $10^{-7} - 10^{-6}$ level, which is in the
sensitivity range of the B-factories.

We use a data sample of $152\times 10^6$ $B\overline{B}$ pairs,
corresponding to an integrated luminosity of 140
fb$^{-1}$, collected by the Belle detector 
at the KEKB~\cite{KEKB} asymmetric energy $e^+e^-$ 
collider. The Belle detector is a large-solid-angle
magnetic spectrometer that consists of a three-layer silicon
vertex detector (SVD), a 50-layer central drift chamber (CDC), an
array of aerogel threshold \v{C}erenkov counters (ACC), a
barrel-like arrangement of time-of-flight scintillation counters
(TOF), and an electromagnetic calorimeter (ECL) comprised of
CsI(Tl) crystals located inside a super-conducting solenoid coil
that provides a 1.5~T magnetic field.  An iron flux-return located
outside of the coil is instrumented to detect $K_L^0$ mesons and
to identify muons.  The detector is described in detail
elsewhere~\cite{Belle}.

To identify the charged tracks, the proton ($L_p$), pion ($L_\pi$)
and kaon ($L_K$) likelihoods are determined from the information
obtained by the tracking system (SVD+CDC) and the hadron
identification system (CDC+ACC+TOF).
Prompt proton candidates must satisfy the requirements of $L_p/(L_p+L_K)> 0.6 $
and $L_p/(L_p+L_{\pi})> 0.6$, and not be associated with the decay
of a $\Lambda$ baryon.
The prompt proton candidates are also required
to satisfy track quality criteria based on the track impact
parameters relative to the interaction point (IP). The deviations
from the IP position are required to be within 0.3 cm in the
transverse ($x$-$y$) plane, and within $\pm$3 cm in the $z$
direction, where the $z$ axis is opposite the direction of the positron beam.
%
Candidate ${\Lambda}$ baryons are reconstructed from two oppositely
charged tracks, one treated as a proton and the other as a pion,
and must have an invariant mass within $5\sigma$ of the
nominal $\Lambda$ mass, as well as
a displaced vertex and flight direction consistent with a
$\Lambda$ originating from the interaction point. To reduce
background, a $L_p/(L_p+L_{\pi})> 0.6$ requirement is applied to
the proton-like track.
%
%
Photon candidates are selected from  the neutral clusters within
the barrel ECL (with polar angle between $33^\circ$ and
$128^\circ$) having energy greater than 500 MeV.
We discard any photon candidate if the invariant mass, in combination with any
              other photon above 30 (200) MeV, is within $\pm 18$ ($\pm
              32$) MeV/$c^2$ of the nominal mass of the $\pi^0$ ($\eta$)
              meson.  The above selection criteria are optimized using
              Monte Carlo (MC) simulated event samples.

Candidate $B$ mesons are formed
by combining a proton with a $\bar{\Lambda}$ and a photon~\cite{conjugate},
 each defined using the above criteria, and requiring
the beam-energy constrained mass, $\mb =
\sqrt{E^2_{\rm beam}-p^2_B}$, and the energy difference, $\de =
E_B - E_{\rm beam}$,
to lie in the ranges 5.2 GeV/$c^2 < \mb < 5.29$
GeV/$c^2$ and $-0.2$ GeV $ < \de < 0.5$ GeV.
              Here, $p_B$ and $E_B$ refer to the momentum and energy,
              respectively, of the reconstructed $B$ meson, and
              $E_{\rm beam}$ refers to the beam energy, all in the
              $\Upsilon$(4S) rest frame.

The dominant background is from continuum $e^+e^- \to q\bar{q}$
processes (where $q = u, d, s, c$) while the background from $B$
decays is negligible except for possible feed-down
events from $B^+ \to \psigg$. This is confirmed with an off-resonance
data set (10 fb$^{-1}$) accumulated at an energy that is 60 MeV
below the $\Upsilon({\rm 4S})$, and an MC sample of 120 million
continuum events. In the $\Upsilon({\rm 4S})$ rest frame,
continuum events are jet-like while $B\overline{B}$ events are
spherical. We follow the scheme defined in Ref.~\cite{etapk} and
combine seven shape variables to form a Fisher
discriminant~\cite{fisher} in order to maximize the distinction
between continuum processes and signal. The variables used have
almost no correlation with $\mb$ and $\de$. Probability density
functions (PDFs) for the Fisher discriminant and the cosine of the
angle between the $B$ flight direction and the beam direction in
the $\Upsilon({\rm 4S})$ frame are combined to form the signal
(background) likelihood ${\cal L}_{\rm s}$ (${\cal L}_{\rm b}$). We require the
likelihood ratio ${\cal R} = {\cal L}_{\rm s}/({\cal L}_{\rm
s}+{\cal L}_{\rm b})$ to be greater than 0.75; this suppresses
about 86\% of the background while retaining 78\% of the signal.
The selection point is determined by optimizing
$N_s/\sqrt{N_s+N_b}$, where $N_s$ and $N_b$ denote the number of
signal and background; here a signal branching fraction of
$4\times10^{-6}$ is assumed.


\begin{figure}[t!]
\epsfig{file=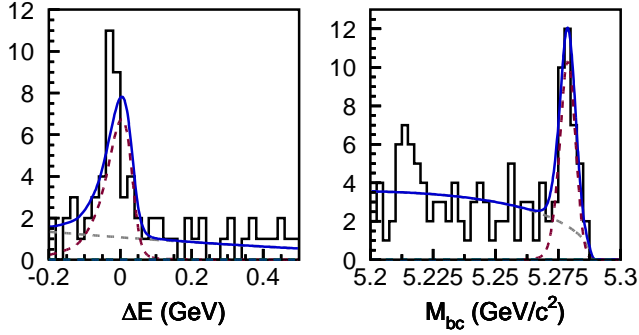,width=3.3in} \caption{The distributions of
$\de$ (for $M_{\rm bc} > 5.27\,{\rm GeV}/c^2$) and $\mb$ (for
$-0.135\,{\rm GeV} < \Delta E < 0.074\,{\rm GeV}$) for $B^0 \to
\plg$ candidates having $M_{p\bar{\Lambda}} < 2.4\,{\rm GeV}/c^2$.
The solid, dotted and dashed lines represent the combined fit
result, fitted background and fitted signal, respectively.}
\label{fg:plgmbde}
\end{figure}

We perform an unbinned extended maximum likelihood fit to the
events with $-0.2$ GeV$<\de<$ 0.5 GeV and $\mb>$ 5.2 GeV/$c^2$ to
determine the signal yields. The extended likelihood function is
defined as
\[\small
 {\cal L} =\frac{e^{-(N_\Lambda+N_\Sigma+N_{q\bar{q}})}}{N!}\prod_{i=1}^{N}
\left[\mathstrut^{\mathstrut}_{\mathstrut}
 N_\Lambda P_\Lambda(M_{{\rm bc}_i},\Delta{E}_i)\right.\]
\[\left.\mathstrut^{\mathstrut}_{\mathstrut}+N_\Sigma
P_\Sigma(M_{{\rm bc}_i},\Delta{E}_i)+
N_{q\bar{q}}P_{q\bar{q}}(M_{{\rm bc}_i},\Delta{E}_i)\right],
\]
where $P_\Lambda$, $P_\Sigma$, and $P_{q\bar{q}}$ are the
PDFs for $\plg$, $\psigg$, and continuum background, respectively, and
$N_\Lambda$, $N_\Sigma$, and $N_{q\bar{q}}$ are the corresponding number
of candidates.

The $\plg$ and $\psigg$ PDFs are
two-dimensional smooth histograms determined by MC simulation.
We use the parametrization first suggested by the ARGUS
collaboration~\cite{Argus}, $ f(\mb)\propto \mb\sqrt{1-(\mb/E_{\rm
beam})^2} \exp[-\xi (1-(\mb/E_{\rm beam})^2)]$, to model the
background $\mb$ distribution and a quadratic polynomial for the
background $\de$ shape. We perform a two-dimensional unbinned fit
to the $\de$ {\it vs} $\mb$ distribution, with the signal and
background normalizations as well as the continuum background
shape parameters allowed to float.


The $\de$ distribution (with $\mb >$ 5.27 GeV/$c^2$) and the $\mb$
distribution (with -0.135 GeV$<\de<$ 0.074 GeV) for the region
$\mpl<$ 2.4 GeV/$c^2$ are shown in Fig.~\ref{fg:plgmbde} along with the
projections of the fit.  
The two-dimensional unbinned fit gives a
$B^+\to\plg$ signal yield of $34.1^{+7.1}_{-6.6}$ with a
statistical significance of $8.6$ standard deviations and a
$B^+\to\psigg$ yield of $0.0\pm4.7$. The significance is
defined as $\sqrt{-2{\rm ln}(L_0/L_{\rm max})}$, where $L_0$ and
$L_{\rm max}$ are the likelihood values returned by the fit with
signal yield fixed at zero and its best fit value,
respectively~\cite{PDG}.

We measure the differential branching fraction of $\plg$ by fitting 
the yield in bins of $\mpl$, as shown in Fig.~\ref{fg:phase}, and
correcting for the corresponding detection efficiency as determined
from a large MC sample of events distributed uniformly in phase space.
The results of the fits along with the efficiencies and the partial
branching fractions are given in Table~\ref{bins}. 
(In these fits, the signal yields are constrained to be non-negative.)
The observed mass distribution in Fig.~\ref{fg:phase} peaks at low
$p\bar{\Lambda}$ mass, a feature seen also in $B^0\to\plpi$ and
$B^+\to p\bar{p}K^+$ decays~\cite{plpi,pph}.

The
photon energy spectrum in the $\Upsilon$(4S) rest frame 
is measured using the same 
fit procedure and is shown in
Fig.~\ref{fg:egamma}. The predicted shape for the generic $b \to s \gamma$ 
process, obtained by MC simulation, is 
shown in Fig.~\ref{fg:egamma} as the shaded histogram. The two distributions
agree with each other within the statistical uncertainty of the present
measurement, although the mean of the $\plg$ spectrum is slightly higher.

\begin{figure}[htb]
\centering \mbox{\epsfig{figure=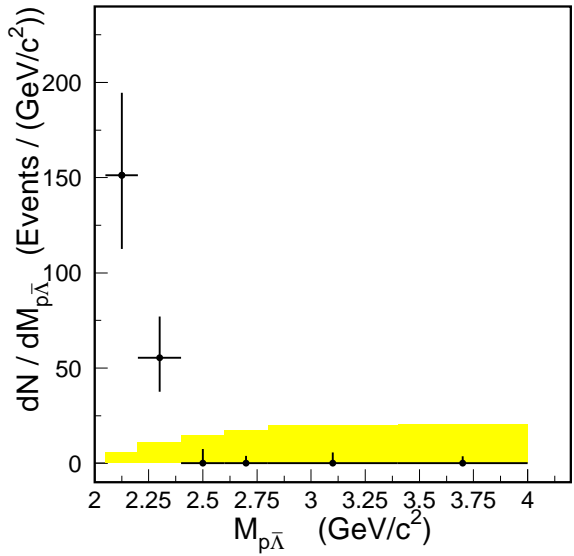,width=3.4in}}
\centering \caption{The differential yield for
$B^0\to \plg$ as a function of $M_{p \bar{\Lambda}}$. The shaded
distribution is from a phase-space MC simulation with area scaled to
the observed signal yield.} \label{fg:phase}
\end{figure}

\begin{table}[htb]
\caption{The event yield, efficiency, and branching
fraction (${\cal B}$)
for each  $M_{p\bar{\Lambda}}$ bin.}
\label{bins}
\begin{center}
\begin{tabular}{cccc}
$M_{p\bar{\Lambda}}$ (GeV/$c^2$)& ~~Signal Yield~~&
~~Efficiency(\%)~~& ${\cal B}$ ($10^{-6}$)
\\
\hline $<2.2$& $22.7^{+6.5}_{-5.8}$& 10.6& $1.41^{+0.40}_{-0.36}$
\\
\hline $2.2-2.4$& $11.1^{+4.3}_{-3.6}$& 9.8&
$0.74^{+0.29}_{-0.24}$
\\
\hline $2.4-2.6$& $0.0^{+1.5}_{-1.5}$& 9.3& $0.00^{+0.11}_{-0.11}$
\\
\hline $2.6-2.8$& $0.0^{+0.8}_{-0.8}$& 9.9& $0.00^{+0.06}_{-0.06}$
\\
\hline $2.8-3.4$& $0.0^{+3.4}_{-3.4}$& 9.6& $0.00^{+0.23}_{-0.23}$
\\
\hline $3.4-4.0$& $0.0^{+2.2}_{-2.2}$& 9.6& $0.00^{+0.15}_{-0.15}$
\\
\hline Total & $33.8^{+9.0}_{-8.1}$& - & $2.16^{+0.58}_{-0.53}$
\\
\end{tabular}
\end{center}
\end{table}

\begin{figure}[htb]
\centering \mbox{\epsfig{figure=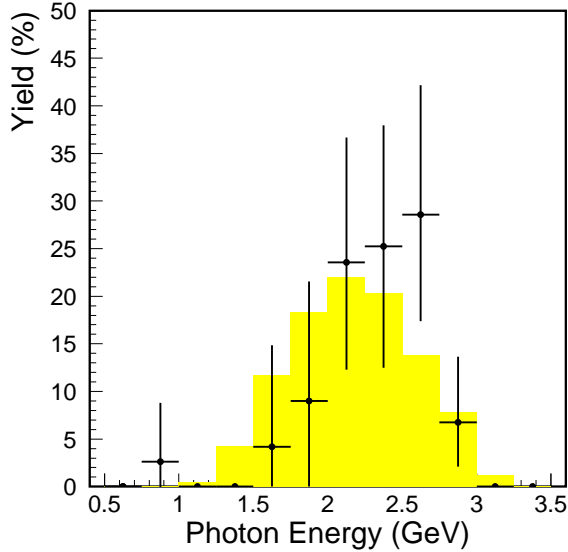,width=3.4in}} \centering
\caption{Signal yield fraction versus photon energy in the
$\Upsilon({\rm 4S})$ rest frame. The shaded histogram is the prediction from
the inclusive $b\to s \gamma$ MC simulation.} \label{fg:egamma}
\end{figure}

We also study the proton angular distribution of the baryon pair
system for $\mpl<4.0$ ${\rm GeV}/c^2$. The angle $\theta_X$ is
measured between the proton direction and the $\gamma$
direction in the baryon pair rest frame. Fig.~\ref{fg:thetap}
shows the efficiency corrected $B$ yield in bins of $\cos
\theta_X$. This distribution confirms the $b \to s \gamma$
fragmentation picture where the $\Lambda$ tends to emerge
opposite the direction of the photon. 

\begin{figure}[htb!]
\epsfig{file=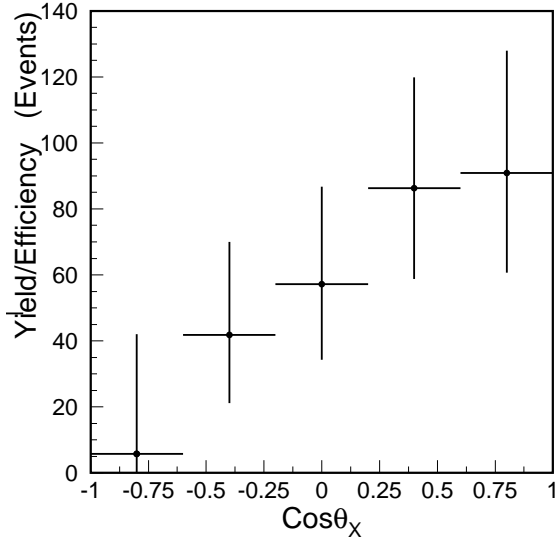,width=3.3in} \caption{Differential branching
fraction versus $\cos\theta_X$ in the baryon
pair system. 
} \label{fg:thetap}
\end{figure}

The systematic uncertainty in particle selection is studied using
high statistics control samples. Proton identification is studied
with a  $\Lambda \to p \pi^-$ sample. The tracking efficiency is
studied with a $D^*$ sample, using both full and partial
reconstruction. Based on these studies, we sum the correlated errors
linearly and assign
a 4.1\% error for proton identification
and 4.9\% for the tracking efficiency.

For $\Lambda$ reconstruction, we have an additional uncertainty of 2.5\% on the
efficiency for off-IP track reconstruction, determined from the
difference of $\Lambda$ proper time distributions for data and MC
simulation. 
There is also a 1.2\%
error associated with the $\Lambda$ mass selection and a $0.5\%$ error for the
$\Lambda$ vertex selection~\cite{2body}. Summing the
errors for $\Lambda$ reconstruction, we obtain
a systematic error of 2.8\%.

The 2.2\% uncertainty for the photon detection is determined from radiative
Bhabha events.  
For the $\pi^0$ and
$\eta$ vetoes, we compare the fit results with and without the vetoes;
the difference of the branching fraction amounts to 0.5\%, and we quote
this as the associated systematic error.

Continuum suppression is studied by changing the selection
criteria on $\cal R$ in the interval 0 -- 0.9 to see if there is
any systematic trend in the signal fit yield. We
quote a 2.5\% error for this.

The systematic uncertainty from fitting is 2.2\%, which is
determined by varying the parameters of
the signal and background PDFs by $\pm 1\sigma$. The MC
statistical uncertainty and modeling with six $\mpl$ bins
contributes a 4.4\% error (obtained by changing the $\mpl$ bin
size). The error on the number
of total $B\overline{B}$ pairs is 0.7\%. 
The error from the sub-decay branching fraction of $\Lambda \to
p\pi^-$ is 0.8\%~\cite{PDG}.

We combine the above uncorrelated errors in quadrature. The total systematic
error is 9.2\%.

We see no evidence for the decay $B^+ \to p\bar{\Sigma}^0\gamma$.
We use the fit results to estimate the expected background, and
compare this  with the observed number of events in the $\psigg$
signal region (-0.20 GeV$<\de<$ 0.04 GeV and $\mb >$ 5.27
GeV/$c^2$) in order to set an upper limit on the
yield~\cite{Highland,Gary,Conrad}. The estimated
background for $\mpl<2.4$ GeV/$c^2$ ($\mpl<4$ GeV/$c^2$) is $45.5
\pm 6.3 (84\pm0.2)$, the number of observed events is 44 (96), and the
systematic uncertainty is 9.2\%; from these, the upper limit yield is
14.9 (35.5) at 90\% confidence level.
The efficiency, estimated from the phase space MC simulation, is
12.1\% (7.0\%), so the 90\% confidence-level upper limits
for the branching fractions are ${\mathcal B}(B^0 \to
\psigg)  < 0.8 \times 10^{-6}$
for ${\mpl<2.4 {\rm GeV/}c^2}$
and
${\mathcal B}(B^0 \to \psigg) < 3.3
\times 10^{-6}$ for ${\mpl<4.0 {\rm GeV}/c^2}$.

In summary, we have performed a search for the radiative baryonic
decays $B^0 \to \plg$, and $\psigg$ with $152$ million $B\bar{B}$
events. A clear signal is seen in the $\plg$ mode, and we measure
a branching fraction of ${\cal B}(B^+ \to \plg)=
{(2.16\,^{+\,0.58}_{-\,0.53} \; ({\rm stat})\pm 0.20 \;
({\rm syst})) \times 10^{-6}}$, which is consistent with the upper
limit set by CLEO\cite{Edwards:2003js}. The yield of the $B^0\to\psigg$ mode
is not statistically significant, and we set the 90\% confidence
level upper limit of
${\mathcal B}(B^0 \to\psigg)  < 3.3 \times 10^{-6}$
for ${\mpl<4.0 {\rm GeV}/c^2}$.


We thank the KEKB group for the excellent operation of the
accelerator, the KEK Cryogenics group for the efficient operation
of the solenoid, and the KEK computer group and the NII for
valuable computing and Super-SINET network support.  We
acknowledge support from MEXT and JSPS (Japan); ARC and DEST
(Australia); NSFC (contract No.~10175071, China); DST (India); the
BK21 program of MOEHRD and the CHEP SRC program of KOSEF (Korea);
KBN (contract No.~2P03B 01324, Poland); MIST (Russia); MESS
(Slovenia); NSC and MOE (Taiwan); and DOE (USA).



\begin{thebibliography}{99}
\bibitem{conjugate} {Throughout this report,
inclusion of charge conjugate mode is always implied
unless otherwise stated.}
\end{thebibliography}
\end{document}